\documentclass{article}

\usepackage{url}
\usepackage{amsmath}
\usepackage{amsfonts}
\usepackage{amssymb}
\usepackage{verbatim}
\usepackage{listings}
  \newcommand{\code}[2][C++]{\lstset{language=#1, basicstyle=\sffamily}\lstinline{#2}}
  \lstnewenvironment{codeblock}[1][C++]
      {\lstset{basicstyle=\small \sffamily}}
      {}

\begin{document}
\title{Functional programming framework for \emph{GRworkbench}}

\author{Andrew J.~Moylan, Susan~M. Scott and Antony~C. Searle\footnote{andrew.moylan@anu.edu.au, susan.scott@anu.edu.au, antony.searle@anu.edu.au} \\
  Centre for Gravitational Physics, \\
  Department of Physics, Faculty of Science, \\
  The Australian National University, \\
  Canberra ACT 0200, Australia. \\
  }

\maketitle

\begin{abstract}
The software tool \emph{GRworkbench} is an ongoing project in visual, numerical General Relativity at The Australian National University.  Recently, the numerical differential geometric engine of \emph{GRworkbench} has been rewritten using functional programming techniques.  By allowing functions to be directly represented as program variables in C++ code, the functional framework enables the mathematical formalism of Differential Geometry to be more closely reflected in \emph{GRworkbench}.  The powerful technique of `automatic differentiation' has replaced numerical differentiation of the metric components, resulting in more accurate derivatives and an order-of-magnitude performance increase for operations relying on differentiation.
\end{abstract}

\section{Introduction}

The goal of the ongoing \emph{GRworkbench} project at The Australian National University is to create a visual software tool for numerical operations on analytically defined space-times in General Relativity.  Such a tool would be used by researchers and educators alike to quickly gain insight into the physical properties of known exact solutions of the Einstein field equation.

A new version of \emph{GRworkbench} was implemented in 1999 \cite{R:grwb-searle-thesis-1999}. It featured a novel numerical differential geometric engine and a flexible visualisation system \cite{R:grwb-grossman-2002, R:grwb-grg-2002}, and was easy to extend with additional space-time definitions.

In 2003, the numerical and differential geometric aspects of \emph{GRworkbench} were rewritten using functional programming techniques, enabling the direct representation in the C++ code of \emph{GRworkbench} of those concepts in numerical computation and Differential Geometry that are normally defined in terms of functions.  The new functional framework enables potentially complex numerical experiments to be quickly and simply defined \cite{R:grwb-grossman-2004}.  Numerical experiments performed in \emph{GRworkbench} were employed in our analysis of a recent scientific claim, described elsewhere \cite{R:grwb-karim-cqg-2004, R:honours-thesis-2003}.\footnote{For the original scientific claim see \cite{R:karim-2003}.}

Functional programming and its applications in \emph{GRworkbench} are described in \S\ref{S:functional}.  Automatic differentiation, in which the derivatives of a function are exactly computed whenever the value of the function is computed, is described in \S\ref{S:automatic-differentiation}.  In \S\ref{S:interval} we outline planned future developments for the numerical engine of \emph{GRworkbench}, in which interval arithmetic will be used to establish a guaranteed bound on numerical errors.

\section{Functional programming}
  \label{S:functional}

In the traditional programming languages of scientific computing, programs typically consist of \emph{routines} that operate on data stored in program \emph{variables}.  Every variable in C++ has a \emph{type}, and there is an approximate correspondence between C++ types and standard mathematical sets.  Table~\ref{T:functional-correspondence} lists some important examples.

\begin{table}
  \begin{center}
    \begin{tabular}{| c | c | l |}
      \hline
      Set & C++ type & Notes \\
      \hline
      $\mathbb{Z}$ & \code{int} & max.\ $\pm (2^{31} - 1)$ \\
      $\mathbb{R}$ & \code{double} & max.\ $\sim \pm 10^{308}$, precision 15 \\
      $\mathbb{R}^n$ & \code{nvector<double>} & (as for \code{double}) \\
      $(A \to B)$ & \code{function<B (A)>} & see Section~\ref{S:functions-as-data} \\
      \hline
    \end{tabular}
    \caption{Correspondence between certain mathematical sets and C++ types in \emph{GRworkbench}.  The term `max.' refers to the largest representable elements of the set; the term `precision' refers to the number of significant figures to which elements of the set are represented.}
    \label{T:functional-correspondence}
\end{center}
\end{table}

The first two sets in Table~\ref{T:functional-correspondence} are, of course, directly representable in some way in every language of scientific computing.  The \code{nvector<T>} type uses the C++ template mechanism (\cite{R:stroustrup-1997}, page 327) to provide a type representing $n$-tuples of any other type \code{T}.  Thus, \code{nvector<double>} represents elements of $\mathbb{R}^n$ and \code{nvector<nvector<double>>} represents elements of $\mathbb{M}_{m \times n}$, the set of $m \times n$ matrices with real-valued entries.

The following is a routine in C++:
\begin{codeblock}
double mean(double a, double b)
{
  return (a + b) / 2;
}
\end{codeblock}
The corresponding mathematical definition is
\begin{equation}
  \label{E:functional-mean}
  \begin{split}
    &\text{mean} \colon \mathbb{R} \times \mathbb{R} \to \mathbb{R}, \\
    &\text{mean} (a, b) = \frac{a + b}{2}.
  \end{split}
\end{equation}
The first line of the routine conveys the same information as the first line of~\eqref{E:functional-mean}: the routine \code{mean} takes two real numbers as arguments, and returns a real number.  The first line of the routine without the routine name or argument names, i.e., \code{double (double, double)}, is the \emph{signature} of the routine; it conveys the same information as $\mathbb{R} \times \mathbb{R} \to \mathbb{R}$ in~\eqref{E:functional-mean}.  The rest of the routine definition, enclosed in braces, encodes the second line of~\eqref{E:functional-mean}.

We may define a \emph{function} as anything which behaves like the routine \code{mean} above, in the sense that it accepts zero or more arguments, and returns a value.  In the most common programming languages of scientific computation, C and Fortran, the \emph{only} possible functions are routines, and so the terms `function' and `routine' are used interchangeably.  The key feature of functional programming is that there can be functions other than the routines typed in by the programmer---functions created while the program is running.  The mechanism to achieve this in C++ is introduced in \S\ref{S:functors}; we first describe how functions can be stored in variables in C++.

\subsection{Functions as data}
  \label{S:functions-as-data}

The capability to store functions in variables is not unique to functional programming.  Most languages used for scientific computation have some way to store a reference to a program routine.  For example, in C and C++ the \emph{address} \code{&f} of a routine \code{double f (double x)} can be stored in a \emph{function pointer} variable of type \code{double (*) (double)}.  Function pointers can be called just like routines.

\emph{GRworkbench} uses the Boost Function library \cite{R:boost-function-2003} to store references to functions.  The Boost Function library provides the templatised type \code{function<T>} representing a function whose signature is \code{T}.  The following code fragment shows how the routine \code{mean} can thus be stored in a variable:\footnote{Anything after the characters \code{//} in a line of code is ignored by the C++ compiler.}
\begin{codeblock}
function<double (double, double)> f = mean;
// the following two lines are now equivalent
double x = f(1, 2);
double x = mean(1, 2);
\end{codeblock}
Observe from the last two lines that the variable \code{f} can be used just like the routine \code{mean}; they are both functions.

In general, if we let $(A_1 \times \cdots \times A_n \to B)$ denote the set of functions from $A_1 \times \cdots \times A_n$ to $B$, then the corresponding C++ type in \emph{GRworkbench} is \code{function<B (A}${}_\text{\code{1}}$\code{, . . . , A}${}_\text{\code{n}}$\code{)>}, where the sets $B, A_1, \ldots, A_n$ correspond to the types $\text{\code{B}}, \text{\code{A}}{}_\text{\code{1}}, \ldots, \text{\code{A}}{}_\text{\code{n}}$.  The fourth row of Table~\ref{T:functional-correspondence} summarises this relationship.

The most important application of the storage of functions in variables is that functions can then be arguments to other functions.  Consider the following routine, which crudely approximates the derivative of a function $f$ at a point $x$:
\begin{codeblock}
double slope(function<double (double)> f, double x)
{
  double h = 0.1;
  return (f(x + h) - f(x - h)) / (2 * h);
}
\end{codeblock}
The corresponding mathematical definition is
\begin{equation}
  \label{E:functional-slope}
  \begin{split}
    &\text{slope} \colon (\mathbb{R} \to \mathbb{R}) \times \mathbb{R} \to \mathbb{R}, \\
    &\text{slope} (f, x) = \frac{f (x + h) - f (x - h)}{2 h}, \quad h = 0.1.
  \end{split}
\end{equation}

Many other numerical algorithms naturally take a function as an argument.  Two examples are
\begin{equation}
  \begin{split}
    &\text{minimise} \colon (\mathbb{R} \to \mathbb{R}) \times \mathbb{R} \to \mathbb{R}, \\
    &\text{minimise} (f, x) = (\text{a local minimum of $f$ near $x$}),
  \end{split}
\end{equation}
and
\begin{equation}
  \begin{split}
    &\text{integrate} \colon (\mathbb{R} \to \mathbb{R}) \times \mathbb{R} \times \mathbb{R} \to \mathbb{R}, \\
    &\text{integrate} (f, a, b) = (\text{numerical estimate of $\int_a^b f(x) \, dx$}).
  \end{split}
\end{equation}

\subsubsection{Creating functions at run-time}
  \label{S:functors}

Consider the following function, defined in terms of the $\text{slope}$ function~\eqref{E:functional-slope}:
\begin{equation}
  \label{E:functional-derivative}
  \begin{split}
    &\text{derivative} \colon (\mathbb{R} \to \mathbb{R}) \to (\mathbb{R} \to \mathbb{R}), \\
    &\text{derivative} (f) = g, \quad g \colon \mathbb{R} \to \mathbb{R}, \quad g (x) = \text{slope} (f, x).
  \end{split}
\end{equation}
For each function $f$, it returns \emph{the function which returns the slope of $f$ at its argument}.

In most traditional languages of scientific computing it is not possible to encode a routine which returns $\text{derivative} (f)$ for all functions $f \colon \mathbb{R} \to \mathbb{R}$.  In C++ it \emph{is} possible to encode this function, by means of a \emph{functor class} (\cite{R:stroustrup-1997}, pages 514--515).  The result is a C++ routine \code{function<double (double)> derivative(function<double (double)> f)}.  (For complete code, including that of the necessary functor class, see \cite{R:honours-thesis-2003}, pages 13--14, 92--94.)

If we were to replace the \code{slope} routine with a more sophisticated algorithm for numerical differentiation, then this \code{derivative} routine would be a good approximation to the mathematical operation of differentiation.  For example, \code{derivative(sin)} would be a good approximation to the function \mbox{\code{cos}.}\footnote{Many standard functions, including \code{sin} and \code{cos}, are built-in to C++.}  Functional programming permits numerical operations, like \code{derivative}, to be expressed in a way which closely resembles the mathematical operations that they approximate.

\subsection{Functional Differential Geometry}

Many fundamental notions in Differential Geometry and General Relativity, such as the action of the metric tensor, and particle world-lines, are functions.  By elevating functions to the same level as traditional data types ($\mathbb{Z}$, $\mathbb{R}$), functional programming makes these notions directly representable as variables in the C++ code of \emph{GRworkbench}.  Table~\ref{T:differential-geometry-correspondence} summarises the correspondence between important concepts in Differential Geometry and their representations in \emph{GRworkbench}.

\begin{table}
  \begin{center}
    \begin{tabular}{| l | l |}
      \hline
      Concept & Representation in \emph{GRworkbench} \\
      \hline
      Coordinates & \code{nvector<double>} \\
      Metric components & \code{nvector<nvector<double>>} \\
      Inter-chart map & see Section~\ref{S:inter-chart-maps} \\
      Point & \code{point} \\
      Tangent vector & \code{tangent_vector} \\
      Metric & {\footnotesize \code{function<double (tangent_vector, tangent_vector)>}} \\
      World-line & \code{function<point (double)>} \\
      \hline
    \end{tabular}
    \caption{Representation of important differential geometric concepts in \emph{GRworkbench}.}
    \label{T:differential-geometry-correspondence}
  \end{center}
\end{table}

\subsubsection{Charts and the metric components}
  \label{S:differential-geometry-charts}

On the $n$-dimensional space-time manifold $\mathcal{M}$, a \emph{chart} is a pair $(\mathcal{U}, \phi)$ representing a coordinate system on the set $\mathcal{U} \subset \mathcal{M}$, where the one-to-one function $\phi \colon \mathcal{U} \to \mathbb{R}^n$ maps points in $\mathcal{U}$ to their coordinates in $\mathbb{R}^n$.  Space-times are defined in \emph{GRworkbench} by the specification of the components of the metric tensor on $\phi (\mathcal{U}) \subset \mathbb{R}^n$ for one or more charts $(\mathcal{U}, \phi)$, and by the definition of the maps $\phi_\beta \circ \phi^{-1}_\alpha$ between the coordinate systems of overlapping pairs of those charts, where $\circ$ denotes function composition.

The coordinates of a point on a chart, $x^i \in \mathbb{R}^n$, are represented by a variable of type \code{nvector<double>}.  The components $g_{ab}$ of the metric tensor at a point on a chart are represented as an $n \times n$ matrix, by a variable of type \code{nvector<nvector<double>>}.  A function which defines the metric components as a function of the chart coordinates $x^i$ might then be of the form
\begin{equation}
  \label{E:differental-geometry-naive-chart}
  \begin{split}
    &\text{chart} \colon \mathbb{R}^n \to \mathbb{M}_{n \times n}, \\
    &\text{chart} (x^i) = g_{ab} |_{x^i},
  \end{split}
\end{equation}
represented in \emph{GRworkbench} by a function of signature \code{nvector<nvector<double>> (nvector<double>)}.  In general, however, the chart coordinates are an open subset of $\mathbb{R}^n$, and so~\eqref{E:differental-geometry-naive-chart} will not be defined everywhere in $\mathbb{R}^n$.  A mechanism is required to represent functions which are only defined on a subset of some other, standard set.  (By `standard set' we mean a set which is already represented by a type in C++, such as those listed in Tables~\ref{T:functional-correspondence} and \ref{T:differential-geometry-correspondence}.)

\emph{GRworkbench} employs the \emph{Boost Optional} library \cite{R:boost-optional-2003} to represent functions which are undefined for some values of their arguments.  The Boost Optional library provides a templatised type \code{optional<T>}, which represents the set $S \cup \{ \varnothing \}$, where $S$ is the set corresponding to the template parameter type \code{T}, and $\varnothing$ is a special value taken by functions at points where they are undefined.

Using the \code{optional} mechanism we rewrite~\eqref{E:differental-geometry-naive-chart} to support charts defined on subsets of $\mathbb{R}^n$:
\begin{equation}
  \label{E:differental-geometry-chart}
  \begin{split}
    &\text{chart} \colon \mathbb{R}^n \to \mathbb{M}_{n \times n} \cup \{ \varnothing \}, \\
    &\text{chart} (x^i) =
    \begin{cases}
        g_{ab} |_{x^i}, &\text{if the $x^i$ are valid chart coordinates;} \\
        \varnothing, &\text{otherwise.}
    \end{cases}
  \end{split}
\end{equation}
The corresponding C++ type is
\begin{equation}
  \label{E:differental-geometry-chart-c++}
  \text{\small \code{function<optional<nvector<nvector<double>>> (nvector<double>)>}},
\end{equation}
for which \emph{GRworkbench} declares the short synonym \code{chart}, using the C++ \code{typedef} mechanism:
\begin{codeblock}
  typedef function<optional<nvector<nvector<double>>> (nvector<double>)> chart;
\end{codeblock}

The \code{optional} mechanism is most useful when the caller of a function cannot know beforehand whether the function will be defined at the arguments to be given to it.  This is the case for callers of functions of the form~\eqref{E:differental-geometry-chart}.  The \code{optional} mechanism thus enables the differential geometric and visualisation algorithms in \emph{GRworkbench} to be coded in such a way that they can operate on any space-time definition, without prior knowledge of the particular coordinate systems (charts) in which they will be working.

\subsubsection{Inter-chart maps}
  \label{S:inter-chart-maps}

For two charts $(\mathcal{U}_\alpha, \phi_\alpha)$ and $(\mathcal{U}_\beta, \phi_\beta)$, the \emph{inter-chart map} from $\phi_\alpha (\mathcal{U}_\alpha)$ to $\phi_\beta (\mathcal{U}_\beta)$ is
\begin{equation}
  \label{E:inter-chart-map}
  \begin{split}
    &\phi_{\alpha\beta} \colon \phi_\alpha (\mathcal{U}_\alpha) \to \phi_\beta (\mathcal{U}_\beta), \\
    &\phi_{\alpha\beta} (x^i) =  (\phi_\beta |_{\mathcal{U}_\alpha} \circ \phi_\alpha^{-1}) (x^i),
  \end{split}
\end{equation}
where $\phi_\beta |_{\mathcal{U}_\alpha}$ is the function $\phi_\beta$ restricted to the set $\mathcal{U}_\alpha$.

The domain of $\phi_{\alpha\beta}$ is, in general, a subset of $\mathbb{R}^n$.  Hence $\phi_{\alpha\beta}$ cannot be represented by a variable of type \code{function<nvector<double> (nvector<double>)>}; instead, the \code{optional} mechanism is again employed.  Thus, an inter-chart map from a chart $(\mathcal{U}_\alpha, \phi_\alpha)$ to a chart $(\mathcal{U}_\beta, \phi_\beta)$ is represented by a function
\begin{equation}
  \label{E:inter-chart-map-final}
  \begin{split}
  &\text{map} \colon \mathbb{R}^n \to \mathbb{R}^n \cup \{ \varnothing \},  \\
  &\text{map} (x^i) =
    \begin{cases}
      (\phi_\beta |_{\mathcal{U}_\alpha} \circ \phi_\alpha^{-1}) (x^i), &\text{if $(x^i) \in \phi_\alpha (\mathcal{U}_\alpha)$ and $\phi_\alpha^{-1} (x^i) \in \mathcal{U}_\beta$;} \\
      \varnothing, &\text{otherwise.}
    \end{cases}
  \end{split}
\end{equation}
The corresponding C++ type is
\begin{equation}
  \label{E:inter-chart-map-grwb}
  \text{\code{function<optional<nvector<double>> (nvector<double>)>}}.
\end{equation}
The C++ \code{typedef} mechanism is used to define a synonym \code{map} for this type.

\subsubsection{World-lines}
\label{S:world-lines}

The abstract notion of a point $p \in \mathcal{M}$, independent of any particular coordinate system, is represented in \emph{GRworkbench} by a C++ type \mbox{\code{point}}.  A \code{point} is constructed from two pieces of information: a \code{chart} which contains it, and its coordinates on that chart.

The \code{operator[ ]} routine of the \code{point} type takes one argument, a variable of type \code{chart}, and returns a variable of type \code{optional<nvector<double>>}, representing the coordinates of the point on the given chart.  (The \code{optional} mechanism is used because a particular point may not have coordinates on the given chart.)  Thus, if \code{p} is a variable of type \code{point}, and \code{c} is a variable of type \code{chart}, then the coordinates of \code{p} on \code{c} are given by \code{p[c]}.  If there is no inter-chart map defined from the point's original chart to the chart \code{c} then $\varnothing$ is returned.

A general curve in space-time, such as a world-line, which may not be defined for all values of its parameter, is a function $\lambda \colon \mathbb{R} \to \mathcal{M} \cup \{ \varnothing \}$; such functions are represented by variables of type  \code{function<optional<point> (double)>}.  Curves in \emph{GRworkbench} must be represented in this form, because they are often defined in terms of numerical algorithms that may not converge everywhere to a solution, even if a solution exists; such a curve returns the value $\varnothing$ for parameter values for which there was no convergence to a solution.  The synonym \code{worldline} is defined for the type \code{function<optional<point> (double)>}.

\subsubsection{Tangent vectors}
\label{S:tangent-vectors}

The abstract notion of a tangent vector $v \in T_p$, where $T_p$ is the tangent space of a point $p \in \mathcal{M}$, is represented in \emph{GRworkbench} by the C++ type \code{tangent_vector}.  A \code{tangent_vector} is constructed from three pieces of information: the \code{point} to whose tangent space it belongs, a \code{chart} containing that point, and the contravariant components of the tangent vector on that chart.

As with the \code{point} type, the \code{operator[ ]} routine of the \code{tangent_vector} type takes one argument, a variable of type \code{chart}, and returns the components of the tangent vector on the given chart, in a variable of type \code{optional<nvector<double>>}.  When the components of a tangent vector are requested on a chart other than that from which the tangent vector was constructed, \emph{GRworkbench} uses the inter-chart map, if it exists, to compute the components.  If $v^i$ are the components of a tangent vector $v$ at a point $p$ on a chart with coordinates $x^i$, then the components on another chart, with coordinates $x^{i'}$, are
\begin{equation}
  \label{E:differential-geometry-vector-transform}
  v^{i'} = \left. \frac{\partial x^{i'}}{\partial x^i} \right|_p v^i = A^{i'}_i v^i.
\end{equation}
The entries of the matrix $A^{i'}_i$ are the derivatives of the inter-chart map $\phi \colon \mathbb{R}^n \to \mathbb{R}^n$ with respect to the coordinates $x^i$ of its argument, evaluated at $p$.  In its most recent version, \emph{GRworkbench} computes $A^{i'}_i$, and thereby the components $v^{i'}$, using the method of \S\ref{S:automatic-differentiation}.

At a point $p$, the metric $g_{ab}$ is naturally considered as the inner product
\begin{equation}
  \label{E:differential-geometry-metric-vectors}
  \begin{split}
    &\text{metric} \colon T_p \times T_p \to \mathbb{R}, \\
    &\text{metric} (u, v) = g_{ab} u^a v^b.
  \end{split}
\end{equation}
This function is encoded in \emph{GRworkbench} in the routine \code{metric}, whose signature is \code{double (tangent_vector, tangent_vector)}.  Also, the \code{operator*} routine of the \code{tangent_vector} type is defined to call \code{metric}, so that if \code{u} and
\\ 
\code{v} are variables of type \code{tangent_vector}, then the expression \code{u * v} is equivalent to the expression \code{metric(u, v)}.  This notation is reminiscent of the two equivalent forms
\begin{equation}
  g_{ab} u^a v^b = u_b v^b
\end{equation}
for the inner product of two tangent vectors.

\subsection{Numerical operations and applications}

We may think of the operation of numerically determining a geodesic by integrating the geodesic equation from initial conditions as a function
\begin{equation}
  \label{E:geodesic-function}
  \begin{split}
    &\text{geodesic} \colon T_p \to (\mathbb{R} \to \mathcal{M} \cup \{ \varnothing \}), \\
    &\text{geodesic} (v) = \lambda, \quad \lambda \colon \mathbb{R} \to \mathcal{M} \cup \{ \varnothing \},
  \end{split}
\end{equation}
where $\lambda$ is the numerically determined geodesic with tangent vector $v$ at $p = \lambda (0)$.  Similarly, the numerical parallel transport of a tangent vector along a curve may be represented by a function
\begin{equation}
  \begin{split}
    &\text{parallel\_transport} \colon (\mathbb{R} \to \mathcal{M} \cup \{ \varnothing \}) \times T (\mathcal{M})
      \to (\mathbb{R} \to T (\mathcal{M}) \cup \{ \varnothing \}), \\
    &\text{parallel\_transport} (f, v) = h, \quad h \colon \mathbb{R} \to T (\mathcal{M}) \cup \{ \varnothing \}, \\
    &\text{$h (t)$ is the parallel transport of $v \in T_{f (0)}$ to $f (t)$ along $f$},
  \end{split}
\end{equation}
or $h (t) = \varnothing$ if, for example, the numerical parallel transport algorithm did not converge to a solution.

In terms of these two functional definitions it is trivial to define the following interesting object:
\begin{equation}
  \begin{split}
    &\text{parallel\_curve} \colon (\mathbb{R} \to \mathcal{M} \cup \{ \varnothing \}) \times T (\mathcal{M})
      \to (\mathbb{R} \to \mathcal{M} \cup \{ \varnothing \}), \\
    &\text{parallel\_curve} (f, v) = \gamma, \quad \gamma \colon \mathbb{R} \to \mathcal{M} \cup \{ \varnothing \}, \\
    &\gamma (t) = \text{geodesic} (\text{parallel\_transport} (f, v) (t)) (1),
  \end{split}
\end{equation}
which represents the world-line of an object which is stationary at a proper distance $\sqrt{g_{ab} v^a v^b}$ with respect to an observer whose world-line is $f$.  By `gluing' together functional objects such as these, it is easy to define potentially complex \emph{numerical experiments} simulating interesting physical situations \cite{R:grwb-karim-cqg-2004}.

\section{Automatic differentiation}
  \label{S:automatic-differentiation}

Differentiation plays a role in most numerical operations in \emph{GRworkbench}, because the derivatives of the components of the metric tensor feature in, for example, the geodesic equation and the parallel transport equation.  Previously in \emph{GRworkbench}, numerical differentiation of arbitrary functions $f \colon \mathbb{R} \to V$, where $V$ is any vector space, was accomplished via numerical estimation of the limit
\begin{equation}
  \label{E:numerical-centred-difference}
  \lim_{h \to 0} \frac{f (x + h) - f (x - h)}{2 h} = \lim_{h \to 0} d (h),
\end{equation}
using the technique of \emph{Richardson extrapolation} (\cite{R:honours-thesis-2003}, pages~18--21; \cite{R:numerical-recipes-1992}, pages~186--189).  In~\eqref{E:numerical-centred-difference}, $d (h)$ is the \emph{centred difference} approximation to the derivative of $f$ at $x$.

While this method is easily applicable to any function, it has some important drawbacks.  Firstly, it requires many evaluations of the function $f$, at $x \pm h$ for numerous values of $h$.  Secondly, an estimate of the limit~\eqref{E:numerical-centred-difference} may not converge, or, what is worse, may converge to a value that is incorrect.  The accurate convergence of the algorithm depends on the values of $h$ at which $d (h)$ is evaluated being of approximately the same size as the smallest scale over which the function varies significantly around $x$.  For a general implementation of the numerical differentiation method~\eqref{E:numerical-centred-difference}, this information may not be available.  Thirdly, the accuracy of this method is rarely more precise than half as many significant figures as the precision of the floating point arithmetic employed.  (For the \code{double} type this is $\sim 15 / 2 \sim 7$ significant figures.)

The technique of \emph{automatic differentiation} (\cite{R:verified-computing-1993}, Chapter~5) avoids all of these difficulties.  The technique follows from three observations.  The first is that all C++ routines encoded by the programmer must, at the lowest level, eventually be defined in terms of a finite set of built-in fundamental operations (such as addition, extraction of square roots, and trigonometric functions).  The second is that all of these fundamental operations have exact derivatives at (almost) every point at which they are defined, and these exact derivatives may themselves be expressed in terms of the fundamental operations.  The third is that from the exact derivatives of the fundamental operations we can obtain the exact derivative of any function defined in terms of the fundamental operations, through the use of the chain rule for differentiation:
\begin{equation}
  \frac{d}{dx} f (u (x)) = \left. \frac{d}{du} f (u) \right|_{u (x)} \frac{d}{dx} u (x).
\end{equation}

These facts are exploited in the method of \emph{forward automatic differentiation}, which is implemented in \emph{GRworkbench}.  In this method, the fundamental C++ operations which act on real numbers are extended to also act on ordered pairs $(u, u')$ of real numbers, with $u$ being the usual argument to the function, and $u'$ representing the derivative of $u$ with respect to some independent variable $x$.  For example, we redefine
\begin{equation}
  \begin{split}
    &\sin \colon \mathbb{R} \times \mathbb{R} \to \mathbb{R} \times \mathbb{R}, \\
    &\sin (u, u') = (\sin u, u' \cos u),
  \end{split}
\end{equation}
and
\begin{equation}
  \begin{split}
    &\text{multiplication} \colon (\mathbb{R} \times \mathbb{R}) \times (\mathbb{R} \times \mathbb{R}) \to \mathbb{R} \times \mathbb{R}, \\
    &(u, u') (v, v') = (u v, u' v + u v').
  \end{split}
\end{equation}

In this way, the derivative of a function is automatically computed whenever the value of the function is computed.  The derivative is accurate up to round-off error in the floating point arithmetic, and its accuracy is independent of the behaviour of the function around $x$.  As well as being more accurate, the method is also faster than numerical differentiation, because the extra computation involved in computing the second part of each ordered pair $(u, u')$ (roughly a factor of two) is generally less computationally expensive than the many evaluations of $f$ that are necessary to numerically estimate the value of the limit~\eqref{E:numerical-centred-difference}.

The replacement of numerical differentiation by automatic differentiation has resulted in a large performance and accuracy increase for those numerical operations in \emph{GRworkbench} that are dependent on the differentiation of the components of the metric tensor.

\section{Future directions}
  \label{S:interval}

Currently in \emph{GRworkbench}, tangent vectors and metrics, respectively tensors of type $(1, 0)$ and type $(0, 2)$, are each implemented separately.  We will replace them with an implementation of general tensors of arbitrary type, whose components on any chart are determined in terms of their components on one chart by the general transformation rules.  This implementation will be based around a C++ type that can represent any function of vectors and covectors that is linear in each of its arguments.  Thus, the representation of tensors in \emph{GRworkbench} will be nearly identical to their usual mathematical definition.

\subsection{Interval arithmetic}

Many computations require us to make
statements not about points, but rather about open sets of
points.  Ordinary differential equation solvers require
that functions be defined, bounded, and have some number of continuous derivatives not at a single point, but in a
neighbourhood around an initial condition.  Just as
we can implement automatic differentiation by extending
the definition of basic operations, we can also implement
\emph{interval arithmetic} (\cite{R:verified-computing-1993}, Chapter~3), so that for
\begin{equation}
f\colon\mathbb{R}\to\mathbb{R}
\end{equation}
we define a new function
\begin{equation}
f'\colon\mathbb{R}\times\mathbb{R}\to\mathbb{R}\times\mathbb{R}
\end{equation}
that satisfies
\begin{equation}
f' (a, b) = (c, d) \Rightarrow \forall \, x\in[a,b], f(x)\in [c, d].
\end{equation}
The function $f'$ is a function from one interval of real numbers to another interval of real numbers.  By combining the definition of a function in terms of fundamental operations with a bound on its parameter, we obtain a bound on the value of the function.  For example, the function
\begin{equation}
  \begin{split}
    &\text{multiply} \colon \mathbb{R} \times \mathbb{R} \to \mathbb{R}, \\
    &\text{multiply} (a, c) = a c,
  \end{split}
\end{equation}
is generalised to act on intervals of real numbers by
\begin{equation}
  \begin{split}
    &\text{multiply}' \colon (\mathbb{R} \times \mathbb{R}) \times (\mathbb{R} \times \mathbb{R}) \to (\mathbb{R} \times \mathbb{R}), \\
    &\text{multiply}' ((a, b), (c, d)) = \left( \min(ac, ad, bc, bd), \max(ac, ad, bc, bd) \right).
  \end{split}
\end{equation}
Thus, the product of the interval $(2.9, 3.1)$ with the interval $(3.9, 4.1)$ is the interval $(11.31, 12.71)$.

When combined with automatic differentiation and the \code{optional} mechanism,
a whole range of statements about \emph{any} function defined in terms of fundamental operations can be
tested on any interval.  We might be able to establish with certainty that a function is everywhere defined
on a certain closed `box' $[a,b]\times[c,d]\times\ldots\subset\mathbb{R}^n$, that it is bounded, or even that it is $C^k$.  These properties
are important to guarantee that geodesic tracing and other algorithms can
respond correctly to singularities and other pathological conditions.

A concrete example occurs for the metric components of the Schwarz\-schild space-time expressed in standard spherical polar coordinates, which are valid separately for $0 < r < 2 M$ and for $2 M < r < \infty$.  Presently in \emph{GRworkbench}, two \code{chart}s are used, respectively covering
$r \in (0, 2M)$ and $r \in (2M, \infty)$, so that a numerical operation, such as geodesic tracing, that does not `know' about the $r = 2 M$ singularity, cannot accidentally `skip over' it by discretely sampling points on either side of it.

Using interval
arithmetic it would, in principle, be possible to employ one `pseudo-\code{chart}' covering the whole range $0 < r < \infty$, because any numerical operation that evaluated the metric components on an interval containing $r = 2 M$ would learn that the metric components are undefined somewhere in that interval, and are unbounded over it---the algorithm should then attempt to investigate a smaller subset of the original interval.

A more widely applicable further benefit of interval arithmetic is
that it can provide powerful control over round-off errors.  The exact result
for any calculation typically lies between two representable floating-point numbers;
round-off error occurs when one of these must be chosen to represent the result, a process
analogous to writing a value to only so many decimal places.  Instead of choosing one
value, an interval arithmetic operation returns both values, which together bound the tightest representable interval enclosing the true result. Interval arithmetic trades the illusory precision of floating point computations for guaranteed accuracy.

Once it is implemented in \emph{GRworkbench}, the `pseudo-algebraic' method of interval arithmetic, which exploits exactly-known properties of the fundamental mathematical operations in C++, will enable a powerful new method of exploring properties of space-times and render the ability to make exact statements about these properties.

\section{Conclusion}

The new functional framework for the differential geometric engine of \emph{GRworkbench} represents the underlying mathematical structure of Differential Geometry more closely than ever before.  Functional programming makes it easier to define systems that model interesting physical situations in numerical experiments.  The technique of automatic differentiation provides accurate and fast derivatives for functions, such as analytically-defined space-time metrics, that are defined in terms of certain fundamental mathematical operations.  In the future, interval arithmetic may enable the numerical engine of \emph{GRworkbench} to be written in such a way that there are no unknown numerical errors at all.

\bibliographystyle{unsrt}
\bibliography{paper}

\begin{thebibliography}{10}

\bibitem{R:grwb-searle-thesis-1999}
A.~C. Searle.
\newblock \emph{GRworkbench}.
\newblock Honours thesis, The Australian National University, 1999.

\bibitem{R:grwb-grossman-2002}
S.~M. Scott, B.~J.~K. Evans, and A.~C. Searle.
\newblock \emph{GRworkbench}: A computational system based on differential
  geometry.
\newblock In V.~G. Gurzadyan, R.~T. Jantzen, and R.~Ruffini, editors, {\em
  Proceedings of the Ninth {Marcel Grossmann} Meeting on General Relativity},
  pages 458--467. World Scientific, 2002.

\bibitem{R:grwb-grg-2002}
B.~J.~K. Evans, S.~M. Scott, and A.~C. Searle.
\newblock Smart geodesic tracing in \emph{GRworkbench}.
\newblock {\em General Relativity and Gravitation}, 34:1675--1684, 2002.

\bibitem{R:grwb-grossman-2004}
A.~Moylan, S.~M. Scott, and A.~C. Searle.
\newblock Developments in \emph{GRworkbench}.
\newblock In {\em Proceedings of the Tenth Marcel Grossmann Meeting on General
  Relativity}, 2005.
\newblock To appear.

\bibitem{R:grwb-karim-cqg-2004}
A.~J. Moylan, S.~M. Scott, and A.~C. Searle.
\newblock Can the {Milky Way} be weighed using earth-based interferometry?
\newblock 2005.
\newblock In preparation.

\bibitem{R:honours-thesis-2003}
A.~Moylan.
\newblock Numerical experimentation within \emph{GRworkbench}.
\newblock Honours thesis, The Australian National University, 2003.

\bibitem{R:karim-2003}
M.~Karim, A.~Tartaglia, and A.~H. Bokhari.
\newblock Weighing the {Milky Way}.
\newblock {\em Classical and Quantum Gravity}, 20:2815--2825, 2003.

\bibitem{R:stroustrup-1997}
B.~Stroustrup.
\newblock {\em The {C++} Programming Language}.
\newblock Addison-Wesley, third edition, 1997.

\bibitem{R:boost-function-2003}
D.~Gregor.
\newblock {\em Boost Function library}, 2003.
\newblock \url{http://www.boost.org/doc/html/function.html}.

\bibitem{R:boost-optional-2003}
F.~Cacciola.
\newblock {\em Boost Optional library}, 2004.
\newblock \url{http://www.boost.org/libs/optional/doc/optional.html}.

\bibitem{R:numerical-recipes-1992}
W.~H. Press, S.~A. Teukolsky, W.~T. Vetterling, and B.~P. Flannery.
\newblock {\em Numerical Recipes in {C}; The Art of Scientific Computing}.
\newblock Cambridge University Press, second edition, 1992.

\bibitem{R:verified-computing-1993}
R.~Hammer, M.~Hocks, U.~Kulisch, and D.~Ratz.
\newblock Numerical toolbox for verified computing {I}.
\newblock In {\em Springer Series in Computational Mathematics}, volume~21.
  Springer-Verlag, 1993.

\end{thebibliography}

\end{document}